\documentclass[]{jkas} 

\def\year{2023} 
\def\volume{56} 
\def\issue{1} 
\def\beginpage{1} 
\def\received{---} 
\def\accepted{---} 
\def\published{---} 
\date{Received \received; Accepted \accepted; Published \published}

\usepackage{graphicx}
\usepackage{txfonts}
\usepackage{lipsum}
 \usepackage{ulem}
\usepackage{lscape}             
\usepackage{placeins}           
\usepackage{xcolor}

\newcommand\halpha{H$\alpha$}

\newcommand\asc{$^{\prime\prime}$}
\newcommand\kmps{km\,s$^{-1}$}

\newcommand\arcsec{$^{\prime\prime}$}

\newcommand\ls{$\lambda{\text -}s$}

\makeatletter
\providecommand{\@LN}[2]{}
\makeatother

\graphicspath{{./}{figures/}}

\title{
Structures and properties of fan-shaped jets: \halpha\ dark blobs and transparent regions
}

\author[1]{Heesu Yang}{0000-0001-5455-2546}
\author[1,2,3]{Maria S. Madjarska}{0000-0001-9806-2485}
\author[1]{EunKyung Lim}{0000-0002-7358-9827}
\author[4,5,6,7]{Daniel N\'obrega-Siverio}{0000-0002-7788-6482}
\author[8]{Klaus Galsgaard}{0000-0001-8882-1708}

\affil[1]{Korea Astronomy and Space Science Institute, 776 Daedeok-daero, Yuseong-gu, Daejeon 34055, Republic of Korea}
\affil[2]{Max Planck Institute for Solar System Research, Justus-von-Liebig-Weg 3, 37077, G\"ottingen, Germany}
\affil[3]{Space Research and Technology Institute, Bulgarian Academy of Sciences, Acad. G. Bonchev Str., Bl. 1, 1113, Sofia, Bulgaria}
\affil[4]{Instituto de Astrof\'isica de Canarias,  E-38205 La Laguna, Tenerife, Spain}
\affil[5]{Universidad de La Laguna, Dept. Astrof\'isica, E-38206 La Laguna, Tenerife, Spain}
\affil[6]{Rosseland Centre for Solar Physics, University of Oslo, PO Box 1029 Blindern, 0315 Oslo, Norway}
\affil[7]{Institute of Theoretical Astrophysics, University of Oslo, PO Box 1029 Blindern, 0315 Oslo, Norway}
\affil[8]{School of Mathematics and Statistics, University of St Andrews, North Haugh, St Andrews, KY16 9SS, Scotland, UK}

\def\corrauthor{%
H. Yang, \email{hsyang@kasi.re.kr}
}

\def\runningauthor{%
Yang et al.
}

\def\runningtitle{%
\halpha~dark blobs and transparent regions in fan-shaped jets
}

\def\keywords{%
Sun:activity --- Sun: atmosphere --- Sun: chromosphere  --- Sun: UV radiation --- Magnetic reconnection
}
\def\abstracttext{%
We investigated three fan-shaped jets observed above sunspot light bridges or nearby sunspot regions.
The study aimed to explore the dynamics and physical properties of jets' features that appear as blob-like structures at the tips of the jets, which we call `dark blobs'. A transparent region is observed beneath the dark blobs, creating a visible gap between the dark blobs and the trailing body of the jets.
These phenomena were studied in chromospheric and transition region imaging and spectral high-resolution co-observations from the Visible Imaging Spectrometer of the Goode Solar Telescope at the Big Bear Solar Observatory and the Interface Region Imaging Spectrograph (IRIS), together with data from the Atmospheric Imaging Assembly (AIA) on board the Solar Dynamics Observatory.
We analyzed the jets' morphology and fine structure. We obtained the spatial scale and the dynamics of the dark blobs that are seen mostly in the wings of the H$\alpha$ line and have a cross-section of about 0.2\asc$-$0.3\asc. The dark blobs and the transparent regions are seen bright (in emission) in the IRIS slit-jaw 1330~\AA, 1400~\AA, and AIA 304~\AA\ images.   The IRIS Si~{\sc iv} 1394~\AA\ spectrum of the brightenings showed blue-shifted emission of about $16\,$\kmps\ with non-thermal velocities of up to $40\,$\kmps.
We also estimated the electron density of the blue-shifted brightenings to be $10^{12.1\pm0.2}$\,cm$^{-3}$.
Our findings likely suggest that we detect the observational signatures of shock waves that generate and/or contribute to the evolution of fan-shaped jets.
}

\begin{document}

\jkashead

\section{Introduction}
\label{sec:intro}

High-resolution observations have revealed various dynamic features along sunspot light bridges (LBs). One of them is the fan-shaped jets. These jets are observed along light bridges (LBs) and polarity inversion lines in active regions \citep[e.g.,][]{shimizu_2009,jiang_2011,robustini_2016,hou_2016,2018ApJ...855L..19R,robustini_2018,bai_2019,pietrow_2022}. They extend to heights up to 200~Mm \citep{li_2016} and occur intermittently and recurrently. Occasionally, the top of the jets exhibits oscillatory behavior \citep{asai_2001,shimizu_2009,robustini_2016,tian_2018,zhang_2017}. They also exhibit horizontal expansion \citep{lim_2020}, resembling the opening of a traditional east-Asian folding fan. 

The fan-shaped jets are seen in absorption in \halpha\ passbands displaying fast plasma motions of about 40--100\,\kmps\ \citep{robustini_2016,li_2016,li_2020}. The body of the jets is seen in emission in ultraviolet (UV) 1330\,\AA~images of the Interface Region Imaging Spectrograph \citep[IRIS;][]{pontieu_2014}, as reported by \cite{jiang_2011}, and was given the name light walls (LWs) by \citet{hou_2022}. Interestingly, only the leading edge of the jets is bright in the IRIS slit-jaw 1400~\AA\ images (SJIs). This is also true for extreme-ultraviolet (EUV) filters such as the 131\,\AA\ \citep[dominated by transition region emission outside flaring regions,][]{2010A&A...521A..21O}, 171\,\AA\ and 304\,\AA\ taken by the Atmospheric Imaging Assembly \citep[AIA;][]{2012SoPh..275...17L} on board the Solar Dynamics Observatory \citep[SDO;][]{pesnel_2012} \citep[see, e.g.,][]{yangs_2015,bharti_2015,zhang_2017,hou_2016,hou_2017,liu_2022}.

Numerous studies have investigated and discussed the potential triggering mechanisms of fan-shaped jets. 
One widely proposed mechanism is sequential magnetic reconnection in the chromosphere \citep[e.g.,][]{shimizu_2009,toriumi_2015,robustini_2016}  between the vertical field lines of the umbra and the emerging magnetic field of the magnetic canopy structure \citep{jurcak_2006,lim_2020,zhao_2022}. \cite{jiang_2011} proposed a model produced from 3D simulations of magnetic reconnection based on a current sheet magnetic configuration with a strong guide field. The jets produced in these simulations display a consequential appearance along the guide field lines, similar to those in observations. The reconnection process in the observations is witnessed by the appearance of transient brightenings along the LBs seen in UV channels (e.g., AIA~1600~\AA) \citep{lim_2020,robustini_2016}. This reconnection process is referred to as ``slipping magnetic reconnection'' \citep{lim_2020}. 

Another candidate for the triggering mechanism of these oscillating fan-shaped jets is the leakage of p-mode waves. This is plausible because leaked p-mode waves are prominently identified in the umbra as umbral oscillations or umbral flashes in LBs. This idea offers a reasonable explanation for the periodicity of less dynamic jets. However, the faster and larger jets propagating at speeds of a few tens of \kmps\ cannot be explained by p-mode oscillations \citep{tian_2018}.

Several possible mechanisms are proposed to explain how the energy deposited in the lower atmosphere by the triggering mechanisms described in the previous paragraphs is converted into kinetic energy (plasma motions) and thermal energy (plasma heating) of the jets.
One of the suggested conversion processes is the slingshot effect, in which the magnetic energy is converted into the kinetic energy of the plasma comprising the jets by the magnetic tension forces of the bent field lines generated by magnetic reconnection \citep{yokoyama_1995,nishizuka_2008}. This idea has been put forward to explain the observational fact of the displacement between the cool \halpha~jets and hot X-ray/EUV jets \citep{canfield_1996,chae_1999}. Another candidate of the intermediate processes is shock waves \citep{suematsu_1982}. Any form of energy deposited in the lower atmosphere causes disturbances and propagates through a series of upward-moving waves. Waves can easily develop into shocks in the stratified low solar atmosphere \citep{chae_2015}. The shock deposits kinetic and thermal energy into the jet plasma. The p-mode leaked waves, or the disturbances generated by magnetic reconnection, would also be a source of the shock waves.

 This study aims to gain further insight into the fine structure, dynamics, and properties of fan-shaped jets, as well as their possible driving mechanisms. We analyze data taken with the Visible Imaging Spectrometer (VIS), a post-focus instrument of the Goode Solar Telescope \citep[GST;][]{cao_2010} at the Big Bear Solar Observatory (BBSO). 
 Fan-shaped jets were observed on 2022 July 11 (hereafter JET1), 2015 September 3 (JET2), and 2016 May 10 (JET3) (see Fig.~\ref{fig:fov}). 
 The fan-shaped jet observed on 2022 July 11, was also simultaneously recorded in data from IRIS. We report the observation of dark blob-like features at the tip of the threads that form fan-shaped jets and a transparent region located below these features, seen in the blue wing of the \halpha\ line (see Fig.~\ref{fig:2022_enlarge}). We explain the two phenomena as the most likely related to shock waves. In Section \ref{sec:obs}, we describe the observation setup and the data reduction procedure. The observational findings are presented in Section \ref{sec:res}, followed by a discussion in Section \ref{sec:disc}. The summary and conclusions are given in Section~\ref{sum_conc}.

\section{Observational material}\label{sec:obs}
VIS is a single Fabry-Perot etalon spectropolarimeter tunable from $5500$ to $7000$\,\AA~with a narrow 0.07\,\AA~bandpass, installed at the focus plane of the 1.6 meter GST at BBSO. It has a field-of-view (FOV) of 70\,\asc$\times$70\,\asc, acquiring diffraction-limited images with the aid of the 308 sub-aperture adaptive optics system \citep{shumko_2014}. 

The observations were taken on 2015 September 3 by scanning the \halpha~line with stepping positions at $\pm1.0,\,\pm0.6,\,0.0$\,\AA, and an exposure time of 7, 12, and 20 ms, respectively. Sixty frames were recorded during each observational burst, and the 25 best frames were saved. The cadence of the dataset is about 30\,s, including a 15\,s delay between each burst. 
The data on 2016 May 10 include images at stepping positions at $\pm 0.8,\,\pm0.4,\,0$\,\AA, with an exposure time of 9, 15, and 20\,ms. Each burst also took 60 frames; the 25 best frames were saved, resulting in a final cadence of about 40\,s, with a 25\,s delay between each burst. 
For the observation performed on 2022 July 11, the \halpha\ line was scanned with a stepping position at $\pm1.0,\,\pm0.8,\,\pm0.6,\,\pm0.4,\,\pm0.2,\,0.0$\,\AA, and exposure times of 7, 9, 12, 15, 17, and 20\,ms, respectively. Again, 60 frames were taken during each burst, of which 25 were saved. The cadence is approximately 20\,s.  VIS data were processed with the speckle reconstruction technique using the Kiepenheuer-Institute Speckle Interferometry Package \citep{woger_2008}. 

During the observation on  2022 July 11, the IRIS performed coordinated observations with BBSO/VIS pointing at active region (AR) 13035. IRIS scanned the area in a very large, dense 320-step raster mode. The raster size is 279\arcsec~$\times$ 175\arcsec. The data also include SJ images in the 1330\,\AA, 1400\,\AA, 2796\,\AA, and 2832\,\AA\ passbands with a raster size of 167\arcsec~$\times$ 175\arcsec. The raster cadence is 2937\,s, and the step cadence is 9\,s with a pixel scale along the slit direction of 0.16\arcsec.

We also used data from AIA/SDO to study the EUV brightenings associated with the \halpha~jets. AIA EUV data have a 12\,s cadence and 0.6\arcsec~$\times$ 0.6\arcsec~pixel size.

The co-alignment between the VIS, the AIA, and IRIS SJI data was achieved by using the shapes of the sunspots observed in the AIA $1700\,$\AA~wavelength images and the IRIS SJ $2796\,$\AA~images. The VIS images remained stable over the observations due to the consistent operation of the adaptive optics system. The alignment accuracy is below 1\arcsec. To determine the rest wavelength for all spectral lines, we fitted the average profile of a quiet region outside the target area with a Gaussian function.

We produce difference images by subtracting the images in the blue wing of the H$\alpha$~line from the red wing images after normalizing each image by its average intensity. Thus, the blue-colored features in these images represent blue-shifted dominated absorption (e.g., jets or upflows).

\section{Results}\label{sec:res}

\begin{figure*}
\centering
\includegraphics[width=0.9\textwidth]{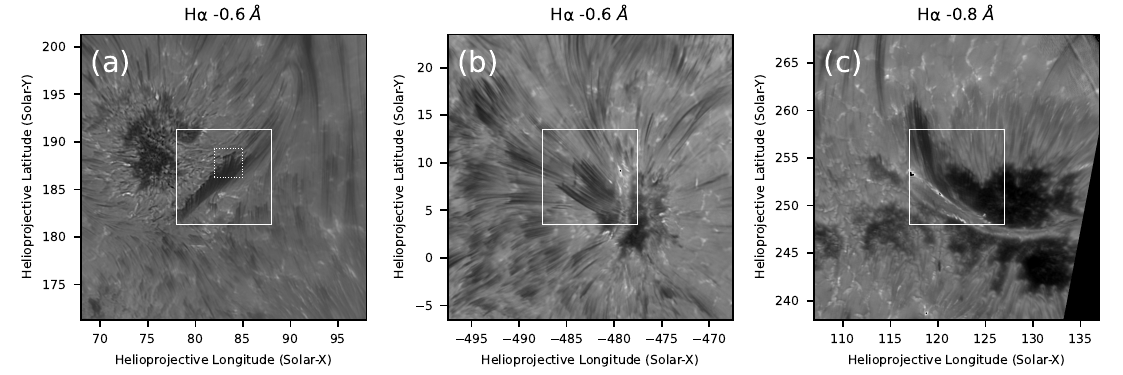}
\caption{Fan-shaped jets. JET1 was observed at \halpha$-$0.6\,\AA~on 2022 July 11 (a), JET2 at \halpha$-$0.6\,\AA~on 2015 September 3  (b), and JET3 at \halpha$-$0.8\,\AA~on 2016 May 10 (c). JET1 (a) appeared near a pore, while JET2 (b) and JET3 (c) originated from LBs.
The solid-line rectangle denotes the FOV shown in Figs.~\ref{fig:2022_case_study}, \ref{fig:150903_aia}, and \ref{fig:160510_aia}. The dotted line square outlines the FOV shown in Fig.~\ref{fig:2022_enlarge}. \label{fig:fov}}
\end{figure*}

Figure \ref{fig:fov} presents snapshots of the three observed fan-shaped jets. Notably, the JET1 shown in Fig.~\ref{fig:fov} (a) is located beside a small sunspot. JET2 and JET3 (Fig.~\ref{fig:fov} (b) and (c), respectively) appear above LBs.

The three jets are composed of vertically elongated threads. When the jets reach their maximum height, the thread tops align, forming flat fronts of the jets. The observed jets are ejected recurrently, and their tops oscillate over time. In the following subsections, we report the investigation of these three jets individually.

\subsection{Jet on 2022 July 11 (JET1)}

\begin{figure*}
\centering
\includegraphics[width=0.9\textwidth]{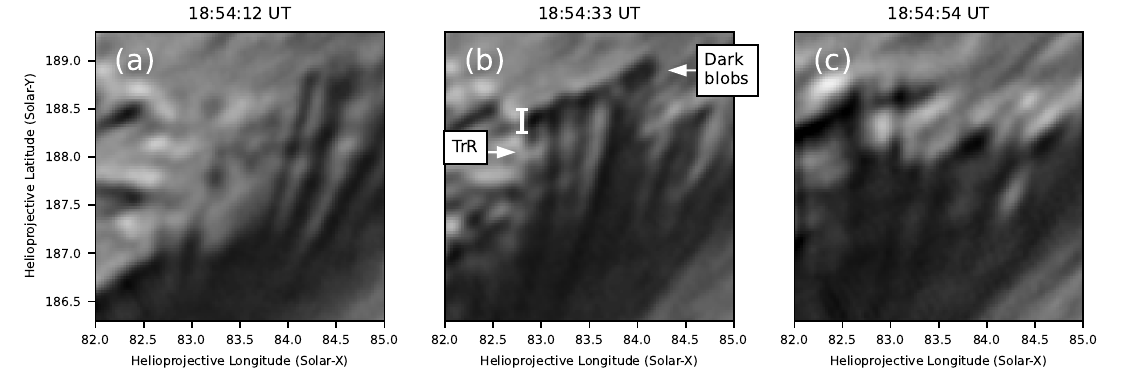}
\caption{Temporal evolution of JET1 seen in series of H$\alpha-$0.6~\AA~images.  The field of view is the area marked with a dotted white-line rectangle in Fig.~\ref{fig:fov} (a). The white tick denotes the average size of the cross-section of dark blobs.}
\label{fig:2022_enlarge}
\end{figure*}

Figure~\ref{fig:2022_enlarge} shows a zoomed view (marked with a white dotted rectangle in Fig.~\ref{fig:fov} (a)) of a temporal sequence of H$\alpha-$0.6 \,\AA~images  of the leading edge of JET1. The leading edge of the jet is indented and appears composed of numerous roundish features in Fig.~\ref{fig:2022_enlarge} (b). We will refer to them as `dark blobs'. The average cross-section of the dark blobs in the plane of the sky is about 0.23\asc~(see a cross-section example shown with a white tick in Fig.~\ref{fig:2022_enlarge}). This value is only twice the diffraction limit of the telescope in the H$\alpha$~band of about 0.1\asc. We note that the cross-section of the blobs is defined as the full-width-half-maximum (FWHM) of the dark absorption structures, determined by fitting a 1D Gaussian function along the direction of the thread of the jet. The length of the indented leading edge of the jet aligning in a parallel direction (vertical to the thread of the jet) is around 1.5\,\asc.

As seen in the H$\alpha$ blue-wing images, the dark blobs appear somewhat separated from the trailing body of the jet. We term this `transparent region' (TrR) marked on panel (b) of Fig.~\ref{fig:2022_enlarge}. The TrR is identified only in the blue wing images \halpha\ (see the next paragraph). In these images, the threads of the jet below the TrR gradually darken towards their base.

\begin{figure*}
\includegraphics[width=\textwidth]{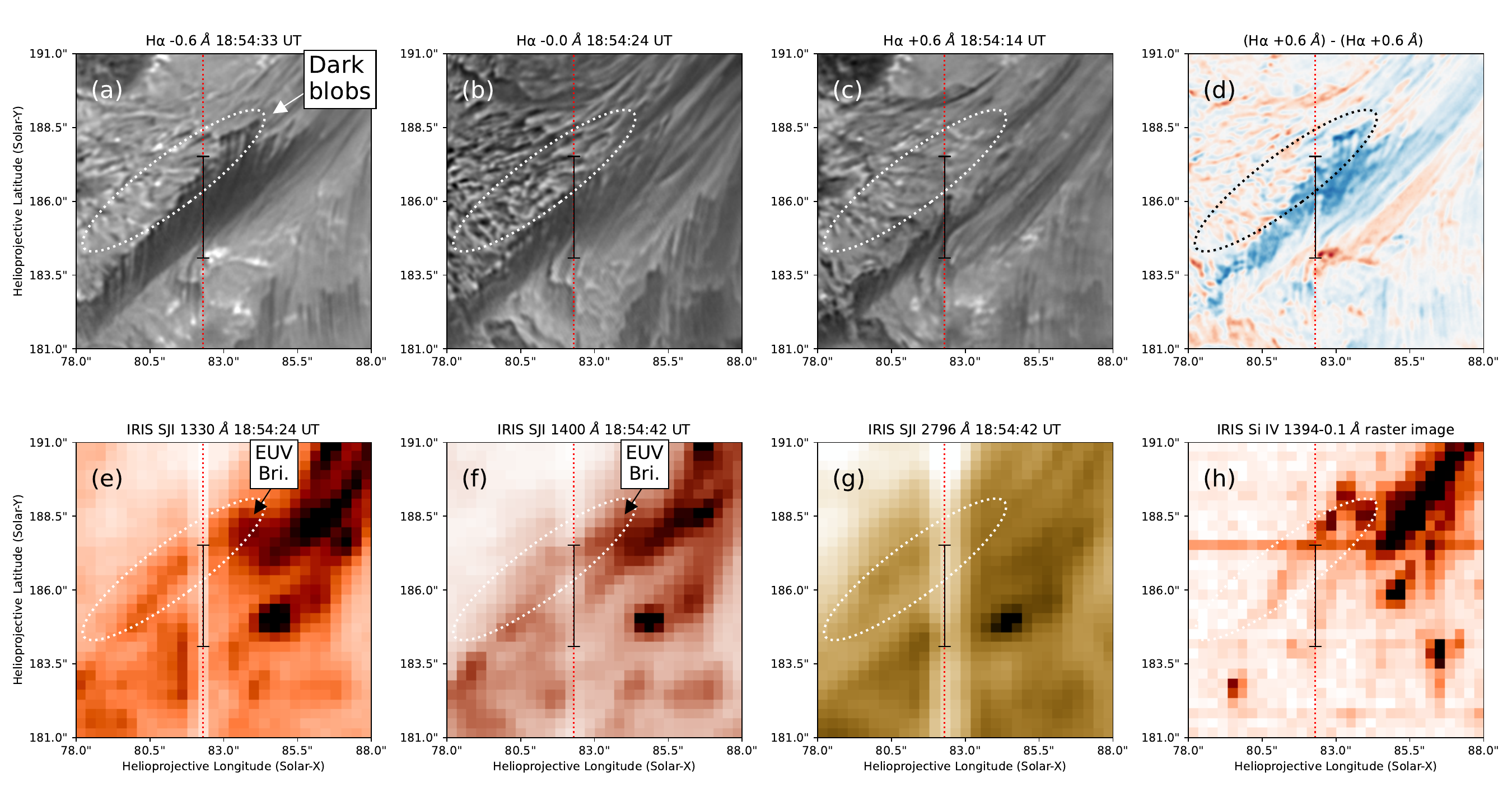}
\caption{JET1 observed near a sunspot. The top row displays the VIS images captured at \halpha$-0.6$\,\AA\ (a), \halpha~center (b), \halpha$+0.6$\,\AA\ (c), and a difference image (see Section~\ref{sec:obs} on the methodology for obtaining this image) (d). The bottom row includes IRIS SJIs (shown as negative) at 1330\,\AA\ (e), 1400\,\AA\ (f), and 2796\,\AA\ (g). Panel (h) features an IRIS raster intensity map in the Si~{\sc iv}~1394$-$0.1\,\AA\ line.
An accompanying animation shows the brightening evolution in the 1330\,\AA\, and 1400\,\AA\ SJIs. All panels share the same field-of-view, corresponding to the white rectangle outlined in Fig.~\ref{fig:fov} (a). The bottom panels are shown as negatives. The red dotted lines mark the IRIS slit position, and the black tick represents the jet extension. The dotted line ellipses mark the location of the brightenings.\label{fig:2022_case_study}}
\end{figure*}


Figure~\ref{fig:2022_case_study} and the accompanying animation show JET1 with its several indented fronts. The tips of the jet threads are seeded with dark blobs seen in the \halpha\ blue-wing images (see panel (a) of Fig.~\ref{fig:2022_case_study}), which are not visible in the \halpha\ line center and red-wing images (see panels (b) and (c), respectively). The difference image in Fig.~\ref{fig:2022_case_study} (d) and the accompanying animation show that the dark blobs and the jet's body are predominantly blue-shifted. 

Panels (e) and (f) of Fig.~\ref{fig:2022_case_study} (shown as negatives), and in detail in Fig.~\ref{fig:2022_enlarge}, reveal that the top of the jet appears in emission in the SJ~1330\,\AA\, and 1400~\AA\ images (hereafter, we will call this region `brightenings'). 
The accompanying animation illustrates the oscillatory behavior of the brightenings, which correlate with the top of the jet visible in the \halpha\ blue wing images. However, due to resolution restrictions, it is not clear whether the brightenings are associated with the TrR or the \halpha\ dark blobs, or both; see the location marked with a white dotted-line ellipse in Fig.~\ref{fig:2022_case_study}. We note that we could not identify the motion along the threads, because the time cadence of our observations is not enough to capture the vertical variations of the dark \halpha~blobs as they also move transversely during their lifetime.

\begin{figure*}[ht!]
\includegraphics[scale=0.75]{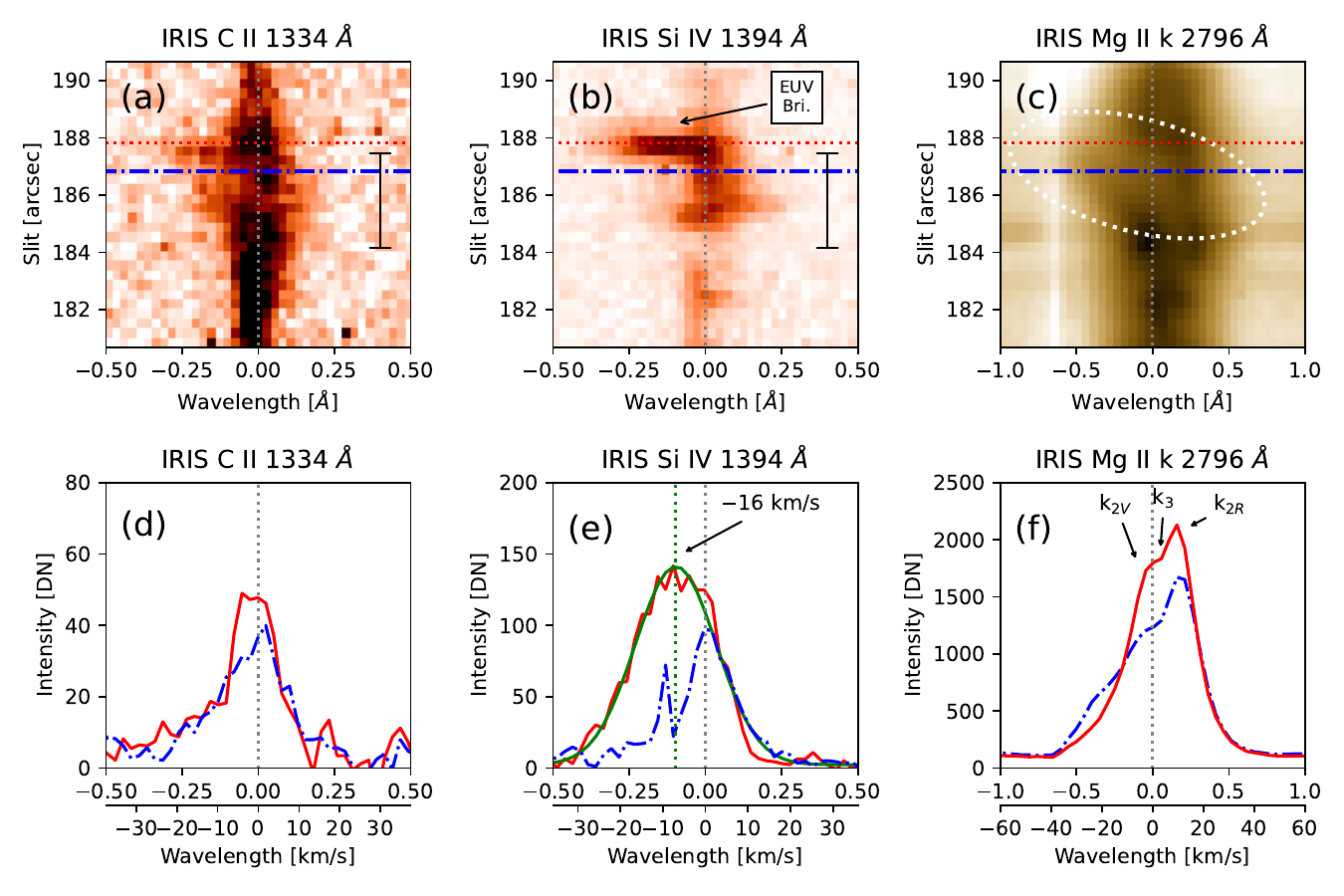}
\caption{JET1 spectral observations. Top panels: Wavelength-space (\ls) plots along the red dotted line in Fig.~\ref{fig:2022_case_study} in C~{\sc ii} 1334\,\AA, Si~{\sc iv} 1394\,\AA, Mg~{\sc ii}~k. The red-dotted and blue-dashed-dotted line profiles in the bottom panel are taken from the location along the slit shown with the corresponding lines in the top panels.  
The green solid line represents the Gaussian fit of the Si~{\sc iv} 1394\,\AA. The large black ticks on panels (a) and (b) mark the jet extension, see also the black ticks in Fig.~\ref{fig:2022_case_study}. The yellow solid line ellipse marks the diagonal absorption feature.}\label{fig:iris_spect_2d}
\end{figure*}

JET1 was also captured under the IRIS slit at 18:54:24~UT. The slit position is marked with a vertical dotted red line in Fig.~\ref{fig:2022_case_study} (e)--(h). Figure~\ref{fig:2022_case_study} (h) shows the raster image in the  Si~{\sc iv}~1394$-$0.1\,\AA\ line. We show the blue wing of the Si~{\sc iv} line because the brightenings of the tips of the jets are enhanced at this wavelength (see Fig.~\ref{fig:iris_spect_2d}). 
The SJ~2796\,\AA~image (Fig.~\ref{fig:2022_case_study} (g)) dominated by the Mg~{\sc ii}~k line reveals the presence of brightenings at the location of the blobs. It also shows that the jet's body is darker than the top of the jet, although the brightenings at the top of the jet are less clear compared to the other SJIs. 

We note that the brightenings at the top of the jets and their oscillations in the SJ 1330\,\AA~and 1400\,\AA\ images have been named `light walls' (LWs) in the studies by \citet{yangs_2015}, \citet{hou_2016}, and \citet{hou_2022}. Specifically, \citet{yangs_2015} suggested that the brightenings are located at the top of the \halpha~LB jets by analyzing \halpha~images from the New Vacuum Solar Telescope at the Yunnan Solar Observatory and IRIS slit-jaw images. While our observations confirm their findings, the present study explores higher-quality images that enable the identification of the jet's fine structures, namely the `dark blobs', which have not been investigated in previous studies.

A strip of pixels with highly enhanced emission is also seen at the bottom of the jet in the SJIs. This bright area appears to be localized along the direction of the jet's lateral expansion 
from northwest to southeast (from the top-right to the bottom-left in the FOV of Fig.~\ref{fig:2022_case_study} (e), (f), and (h)). Assuming the structure originates from the low chromosphere, this area may be the observational signature of magnetic reconnection (see the section~\ref{sec:intro} for more details). Examining the HMI magnetograms, the magnetic configuration at the base of JET1 shows a strong positive polarity emerging between regions of weak negative polarity, which may resemble a light bridge structure. However, this feature is not a subject of interest in the present study and will not be further discussed. We also note that the brightening on the western side of the jet is also identified in Figs.~\ref{fig:2022_case_study} (e) and (f). However, the viewing angle of the structures is not simple, so we do not investigate it further.

At 18:54:33~UT, the IRIS slit is positioned along one of the threads of the jet when the blobs appear. Figure~\ref{fig:iris_spect_2d} displays the wavelength-space (\ls) plots along the slit position marked by a red dotted line in Fig.~\ref{fig:2022_case_study}.  At a latitude of $\approx$188\asc, a strong blue-shifted component is observed in the Si~{\sc iv} and C~{\sc ii} lines. We fitted the top brightening in the Si~{\sc iv} line profile with a single Gaussian to extract the Doppler velocity and the FWHM of the component, as shown by the solid green line in Fig.~\ref{fig:iris_spect_2d} (e). The Doppler velocity of the emission component reaches about $-16$\,\kmps~in the Si~{\sc iv} line. The measured FWHM is 0.31\,\AA~, which corresponds to the non-thermal velocity of the plasma of $40$\,\kmps, by assuming a sharp contribution function peak of the Si~{\sc iv} lines at T = 10$^{4.9}$\,K and the instrumental broadening of $0.026$\,\AA. This is much larger than the known non-thermal speed of about $20$\,\kmps\ in the low transition region \citep{depontieu_2015,cho_2023}.

The Mg~{\sc ii} k line is typically characterized by a central absorption core surrounded by two emission peaks. The k$_3$ denotes the line core of the Mg~{\sc ii} line, and k$_2$, the peaks on both sides of the lines' core. The subscripts V and R mark the blue and red sides of the peaks. Here, the absorption feature from the cool plasma of the body of the jet and the dark blobs is also seen as a decrease in the k$_{2V}$, as shown in Fig.~\ref{fig:iris_spect_2d} (c), marked with a yellow-solid line ellipse. This feature creates a `diagonal' appearance along the slit  (the ellipse inclination marks the diagonal shape). It has also been reported in the cool plasmas of surges observed in \halpha~or Ca~{\sc ii}~8542\,\AA~spectra and was interpreted as shock waves \citep{depontieu_2007,yang_2014}.

We also estimated the electron number density ($N_e$) in the brightenings from the intensity ratios of the O IV 1401\,\AA~and Si IV 1403\,\AA~lines in several pixels during the slit scanning. The methodology is as described in Appendix A. We obtained $N_e$ of $10^{12.1\pm0.2}$\,cm$^{-3}$. This value is similar to the upper range of $N_e$ measured in a surge using the ratio of IRIS O~{\sc iv} 1400~\AA\ and 1401~\AA~lines \citep{nobrega_2021} or in spicules of the order of $10^{12}$\,cm$^{-3}$ \citep{sterling_2000}. However, this value is two orders of magnitude higher than the hydrogen density of the transition region-chromosphere boundary of around $10^{10}$\,cm$^{-3}$ \citep{vernazza_1981,kontar_2008}.

We caution that the electron density obtained using the  O~{\sc iv} 1401\,\AA~and Si~{\sc iv}~1403\,\AA~line ratio yields values of up to an order of magnitude higher than those obtained from the same ion ratios, e.g., O~{\sc iv} line ratios; for details, see the discussions of \citet{judge_2015} and \citet{young_2018}. The low signal-to-noise ratio of the other O~{\sc iv} lines did not allow their use in this study.

\subsection{Jet on 2015 September 3 (JET2)}

\begin{figure*}\centering

\includegraphics[width=1.0\textwidth]{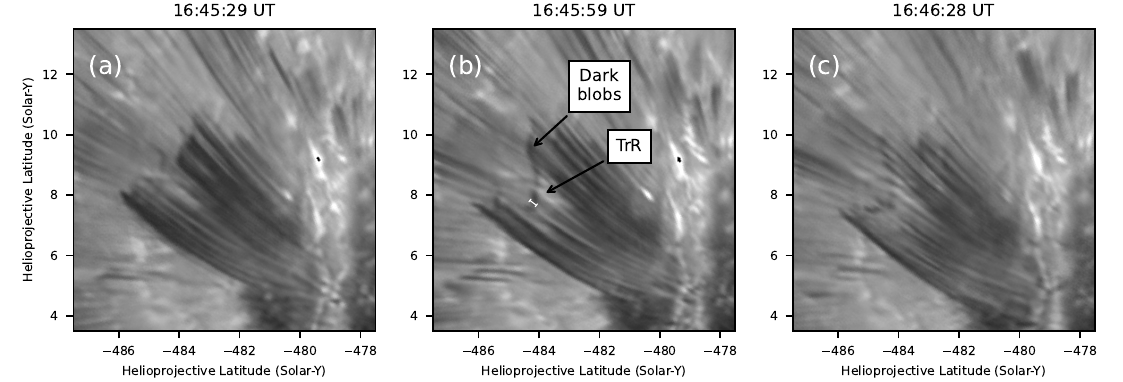}

\caption{Sequential \halpha$-0.6$\,\AA\ images of the blobs observed at around 16:45 UT during JET2. The white tick represents the average cross-section of the \halpha~dark knots. }
\label{fig:150903_darkknots}
\end{figure*}

\begin{figure*}
\centering
\includegraphics[width=0.9\textwidth]{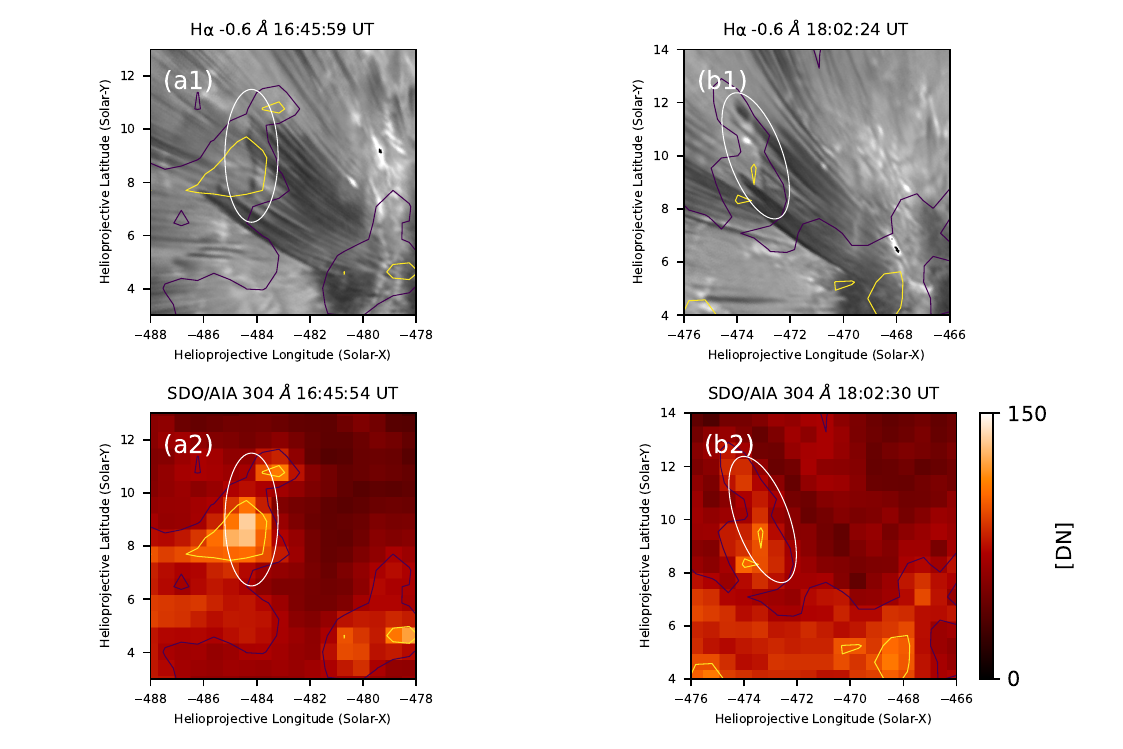}
\caption{Dark blobs seen in \halpha$-$0.6\,\AA\ during JET2 observed at around 16:45~UT (a1). 
 The dark blobs observed at around 18:02~UT  are given in (b1). The AIA 304\,\AA~brightenings are identified at the top of the jets in both events, with contours marking them in (a2) and (b2). White ellipses indicate the blobs identified in (a1) and (b1).}
\label{fig:150903_aia}
\end{figure*}
The light bridge where the JET2 occurs is located in the center of a fully matured sunspot of the active region (AR) NOAA 12409, as shown in Fig.~\ref{fig:fov} (b).  Two sunspots emerged on September~1 and submerged on September~6 within this AR. The light bridge formed in the middle of the two sunspots and persisted over time. The entire process, from formation to destruction of the sunspot, was observed on the visible disk.

We observed intermittent jet threads from the LB. The threads were ejected consecutively, but their height varied. Throughout the observational period, the jets were ejected with intervals of approximately 5$-$7\,min and reached lengths of up to about 7\asc\ in the plane of the sky.

A bundle of taller jet threads was recorded around 16:40~UT. Figure~\ref{fig:150903_darkknots} displays the evolution of the jet ejected around 16:46:28~UT. When the threads reached their maximum height, the faint structures of the \halpha\ dark blobs at the leading edge of the jet threads could be seen in Fig.~\ref{fig:150903_darkknots} (a). In the next snapshot, (b), the dark blobs are identified at a lower height with respect to the previous position recorded in (a). They appear to stretch vertically in the next snapshot (Fig.~\ref{fig:150903_darkknots} (c)). The length of the white tick overplotted on panel (b) of Fig.~\ref{fig:150903_darkknots} represents the cross-section of the blobs. The average cross-section of the blobs along the direction of the jet propagation is about 0.28\asc. A TrR is also clearly identified beneath the blobs in Fig.~\ref{fig:150903_darkknots} (b) and (c). The boundary between the dark blobs and the TrR is quite sharp. However, the boundary between the TrR and the trailing body of the jet is not well-defined, with transparency gradually decreasing towards the bottom of the jet.

The dark blobs are identified in the blue-wing images. The data cadence is approximately 30\,s, which is insufficient to record any motion of the dark blobs, as some of them disappear within a single frame. Therefore, their lifetime is shorter than 30\,s.

Figure~\ref{fig:150903_aia} and the accompanying animations show the evolution of two fan-shaped jets in the \halpha$-$0.6\,\AA~and the corresponding AIA 304\,\AA\ images at a time when the blobs are clearly identified, as marked with white ellipses. The AIA 304\,\AA\ brightenings are localized at the top of the jets. Again, it is unclear whether the brightenings are associated with the blobs or the TrR because of the lower spatial resolution of AIA. An interesting observational fact is that the blobs appear when the AIA~304\,\AA\ brightenings are enhanced (see the accompanying animations). 
The brightenings in AIA~171\,\AA\ (not shown here) are not clearly identified, possibly because the plasma temperature only reaches around $10^5$\,K most of the time, but does not exceed 10$^6$\,K \citep{cheung_2015,antolin_2024}.

\subsection{Jet on 2016 May 10 (JET3)}

\begin{figure*}\centering
\includegraphics[width=0.9\textwidth]{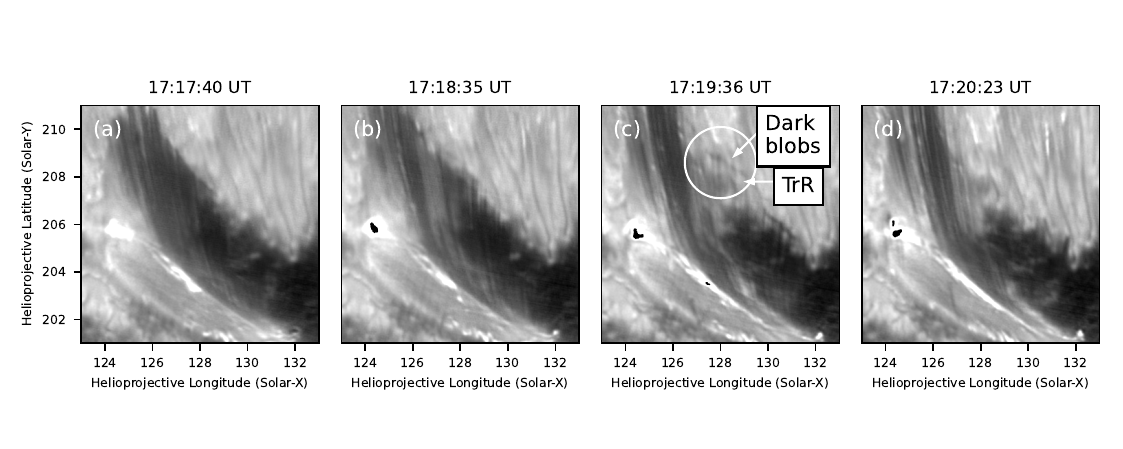}
\caption{Sequential snapshots of JET3 when dark blobs appear at the top in \halpha$-$0.8\,\AA~images on May 10, 2016. The TrR is clearly identified beneath the dark blobs. All panels show the field of view corresponding to the white rectangle outlined in Fig.~\ref{fig:fov} (c).}
\label{fig:160510_darkknots}
\end{figure*}

JET3 appeared quasi-periodically along an LB and has already been studied by \citet{lim_2020} and \citet{yang_2019}, including its origin and magnetic topology. Therefore, we only analyze its leading edge here. We find clear patterns of dark blobs in the \halpha~blue-wing images. Figure~\ref{fig:160510_darkknots} depicts the sequential evolution of the dark blobs observed in the images. The jet extends upward, as seen in  Fig.~\ref{fig:160510_darkknots} (a) and (b). Just after the jet reaches its maximum height (see Fig.~\ref{fig:160510_darkknots} (c)), the area beneath the blobs becomes transparent. The dark blobs disappear in the next frame (see Fig.~\ref{fig:160510_darkknots} (d)),  only $50\,$s after they appeared. 

\begin{figure*}
\centering
\includegraphics[width=0.9\textwidth]{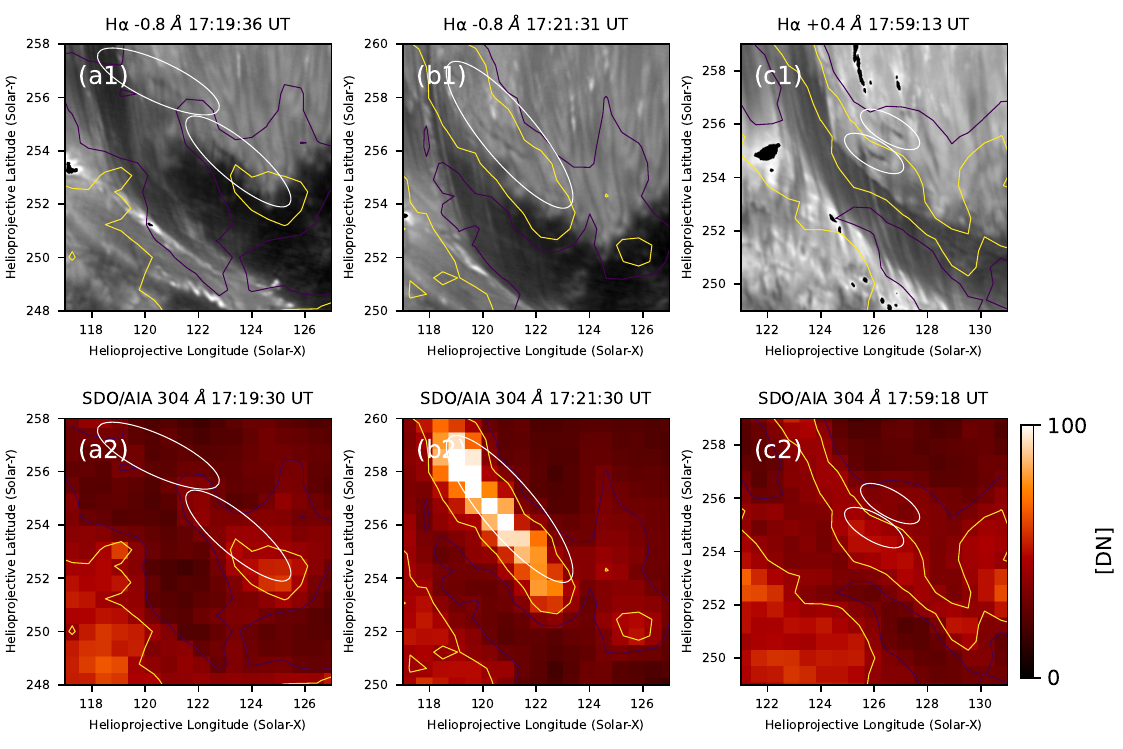}
\caption{Snapshots of dark blobs in \halpha$-$0.8\,\AA~observed at around 17:19 UT on 2016 May 10 (a1). In (a2)  the corresponding AIA~304\,\AA~images are displayed. In (b1), dark blobs are observed at around 17:21 UT in \halpha$-$0.8\,\AA~and in (b2), the corresponding 304\,\AA\ are shown. In (c1), the dark blobs at around 17:59 UT are observed in \halpha+0.4\,\AA, and (c2) shows the corresponding AIA~304\,\AA\ images. The contours mark the AIA~304\,\AA\ brightenings at the edge of the jet, and white ellipses mark the dark blobs identified in the \halpha~images. }
\label{fig:160510_aia}
\end{figure*}

Panels (a1), (b1), and (c1) of Fig.~\ref{fig:160510_aia}  present three snapshots when the dark blobs are clearly identifiable. One can easily distinguish the AIA~304\,\AA~brightenings occurring around the \halpha~dark blobs (panels (a2), (b2), (c2) and accompanying animations), with the brightenings most evident in panels (b2) and (c2). 
We note that the dark blobs in this jet are also seen in the red wing of the \halpha\ line (see panel (c1)). This could be related to the temporal evolution of the dark blobs, as this jet was observed at the highest cadence of 20 s. Data at a higher cadence are needed to further investigate the red wing appearance of the dark blobs. 

\section{Discussion}\label{sec:disc}

\begin{figure}
\includegraphics[width=0.5\textwidth]{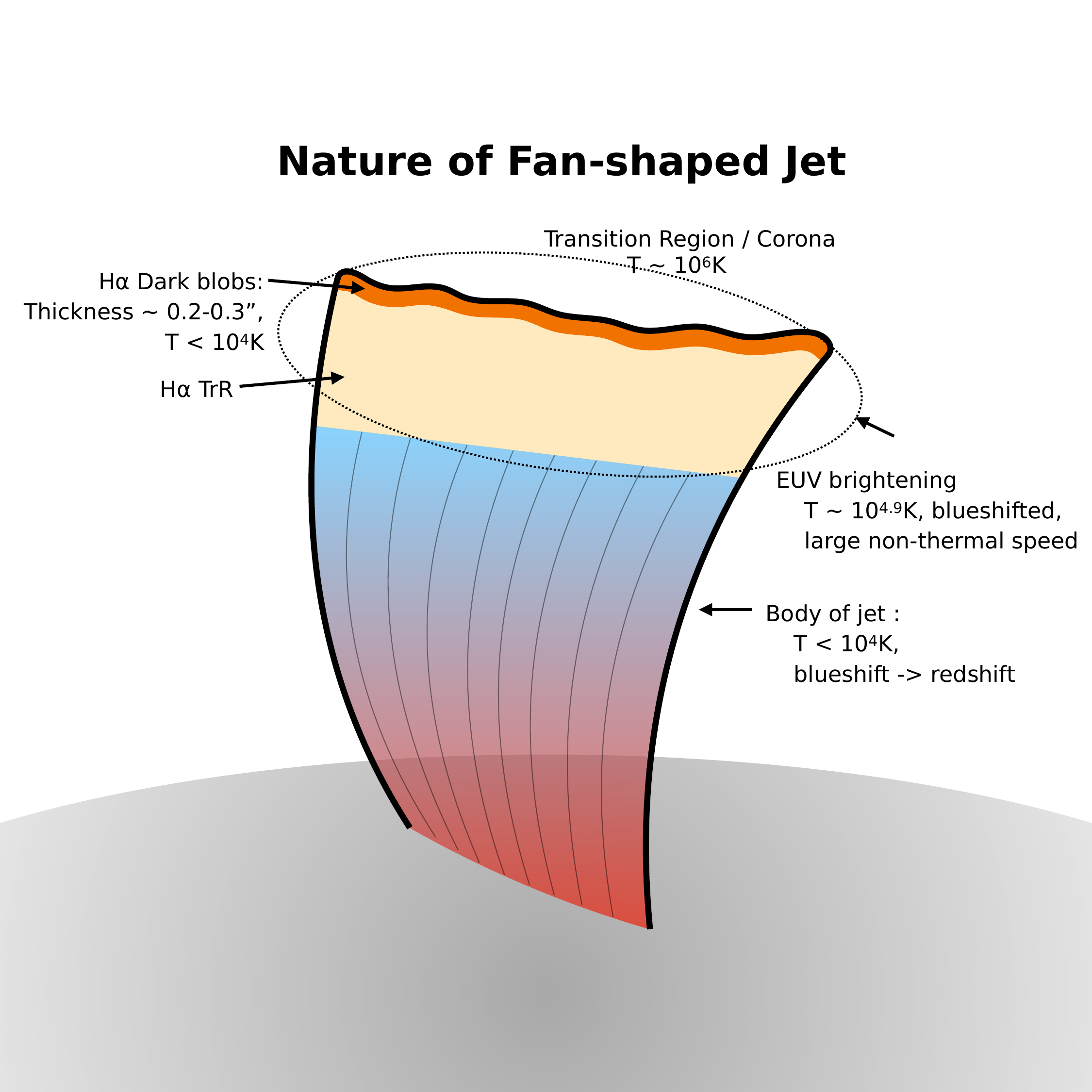}
\caption{Schematic illustration of a fan-shaped jet ejected from the sunspot light bridge, including the observational phenomena newly reported in this paper.}
\label{fig:model}
\end{figure}

We illustrate schematically our observational results in Fig.~\ref{fig:model}. The jets can be divided into the following regions: the dark blobs, the TrR, and the trailing body of the jet beneath the TrR. 
The brightenings identified in the far UV SJ~1330 and 1400~\AA, and EUV AIA~304~\AA\ images (hereafter UV/EUV images), are co-spatial with both the dark blobs and TrR in the \halpha\ images. The lower spatial resolution of the IRIS instrument prevents us from specifying which of the two regions is co-located with the UV/EUV brightening.

The transient brightenings identified in UV/EUV images at the leading edge of the jets suggest that upward-propagating shock waves along fan-shaped jets are most probably responsible for the observed phenomena, as discussed in previous studies \citep{morton_2012,zhang_2017}. The dark \halpha\ blobs and the TrRs co-locating with the UV/EUV brightenings may represent the footprints of the exact location of the shock front. If the UV/EUV brightening is taken to be located at the TrR, the transparency and its variations along the thread in the \halpha~images can be reasonably explained by the plasma temperature reaching around $10^5$\,K where hydrogen is fully ionized and therefore, the region appears in emission in the Si~{\sc iv} lines. In this case, the boundary between the TrRs and the dark \halpha~blobs would correspond to the shock front, exhibiting a temperature jump from approximately $10^4$\,K to $10^5$\,K. The large non-thermal velocity of the blueshifted Si~{\sc iv} component can be interpreted as turbulence in the shocked gas. The high electron density reaching $10^{12.1\pm0.2}$\,cm$^{-3}$ could also be interpreted as an enhancement by the compression and/or the abrupt ionization at the shock front. The increase in opacity in \halpha~images in the lower part of the TrR can then be related to the cooling and recombination that occur behind the shock. If this is the case, the question arises, namely, where does the cool plasma at $10^4$~K (seen as dark blobs) in front of the shock originate? 
One possible explanation is that the boundary between the TrRs and the dark \halpha~blobs marks a transition layer where the wave speed changes from supersonic to subsonic. When plasma ejections in the jet occur successively \citep{suematsu_1982, yang_2014}, the shock waves propagating along the previously ejected plasma dissipate into normal waves that no longer heat the plasma above the transition layer. Then the remaining cool plasma above the layer can be identified as the dark \halpha\ blobs.

If the transient brightening identified in the UV/EUV images is co-located with the dark \halpha~blobs, the mixture of hot and cool plasma must exist in the same spatial location, resulting from a non-equilibrium state following the shock, as suggested by \citet{morton_2012}. In this case, the length of the dark \halpha\ blobs along the jet direction may represent the relaxation time of the shocked gas. Assuming a typical chromospheric sound speed of 10\,\kmps, the cross-sections of the dark \halpha~blobs of 0.2$-$0.3\asc~correspond to a relaxation time of approximately 15$-$20\,s. We note that if this explanation holds, the $N_e$ value derived from the line intensity ratio will not be valid, as one of the fundamental assumptions of the used method is that the plasma is in a steady state.

Why are patterns explained using shock waves more readily identified in fan-shaped jets? This is because the structure of fan-shaped jets provides preferable conditions for generating shock waves compared to other active region jets or surges. Magnetic canopy structures above the LBs can increase the wave amplitude by collimating the wave energy in a funnel. The strong magnetic field in and around a light bridge also suppresses the growth of instabilities, preventing the dissipation of the shocks. Otherwise, shock waves can be easily dissipated. 

Since our observations are limited by the spatial resolution of IRIS and the temporal resolution of VIS, the detailed mechanism occurring within the jets put forward in this study is rather speculative. It will help, however, to validate future modeling. It also highlights the need for a more comprehensive understanding by further exploring observations with comparable high-resolution imaging spectroscopy in both EUV and visible wavelengths, as well as shorter time cadences.

\section{Summary and conclusions}\label{sum_conc}

Here, we investigated a phenomenon in three fan-shaped jets named JET1, JET2, and JET3. Our observational results can be summarized as follows:
\begin{itemize}
\item We observed \halpha~dark blobs located at the leading edges of fan-shaped jets that are visible in absorption mostly in the \halpha\ blue-wing images. The typical cross-section of the dark blobs in the direction of the jet thread is about $0.2-0.3$\,\asc. They are short-lived, so we could not track their variations over time in the VIS data, which has a cadence of $20-60$\,s.
\item Beneath the dark blobs, we identified a transparent region (TrR) seen in the \halpha~images. The boundary between the dark blobs and the TrR is well-defined. The TrR's opacity increases gradually towards the bottom of the jet.
\item The dark blobs and TrRs are associated with UV/EUV brightenings seen in the IRIS/SJ 1330\,\AA, 1400\,\AA, and SDO/AIA 304\,\AA~images for all three events. One of the jets, JET1, for which spectroscopic data are available,  exhibits blue-shifted emission and has large non-thermal velocities.
\item We interpret the formation of the dark \halpha\ blobs and TrR as resulting from the propagation of shock waves and propose models that may explain our interpretation.

\end{itemize}

To gain a better understanding of the dynamics and physical properties of the newly identified feature of fan-shaped jets, specifically dark blobs and adjacent TrRs, observations of chromospheric spectra with a time cadence of less than a few seconds and high spatial resolution ($<$0.1\arcsec) are necessary. Additionally, a higher spatial resolution of UV/EUV images is crucial for resolving the shock heating process at the front of the jets in full detail. For instance, at present, the Extreme Ultraviolet Imager \citep[EUI;][]{Rochus_etal:2020} on board the Solar Orbiter \citep{Muller_etal:2020} together with the Daniel K. Inouye Solar Telescope \citep[DKIST;][]{Rimmele_etal:2020}, or other ground-based data of similar or close resolution, could be employed to investigate these phenomena further.


\acknowledgments
This work was supported by the National Research Foundation of Korea(NRF) grant funded by the Korea government(MSIT) (RS-2022-NR071796). M.M. acknowledges the support of the Brain Pool program funded by the Ministry of Science and ICT through the National Research Foundation of Korea (RS-2024-00408396) and DFG grant WI 3211/8-2, project number 452856778. E.-K.L. was supported by the Korea Astronomy and Space Science Institute under the R\&D program of the Korean government (MSIT; No. 2025-1-850-02). D.N.S. acknowledges support by the European Research Council through the Synergy Grant number 810218 (``The Whole Sun'', ERC-2018-SyG) as well as by the International Space Science Institute (ISSI) in Bern, through ISSI International Team project \#535 Unraveling surges: a joint perspective from numerical models, observations, and machine learning. IRIS is a NASA small explorer mission developed and operated by LMSAL with mission operations executed at NASA Ames Research center and major contributions to downlink communications funded by ESA and the Norwegian Space Centre.
BBSO operation is supported by a US NSF AGS 2309939 grant and the New Jersey Institute of Technology. The GST operation is partly supported by the Korea Astronomy and Space Science Institute and the Seoul National University. 
SDO is a mission for NASA's LWS program. AIA is an instrument on board the Solar Dynamics Observatory. All SDO data used in this work are available from the Joint Science Operations Center \href{http://jsoc.stanford.edu}{http://jsoc.stanford.edu} without restriction.


\appendix
\section{ Diagnosing the electron number density of the UV/EUV brightening at the edge of the jet.}\label{section:appendixA}

\begin{figure*}
\includegraphics[width=\textwidth]{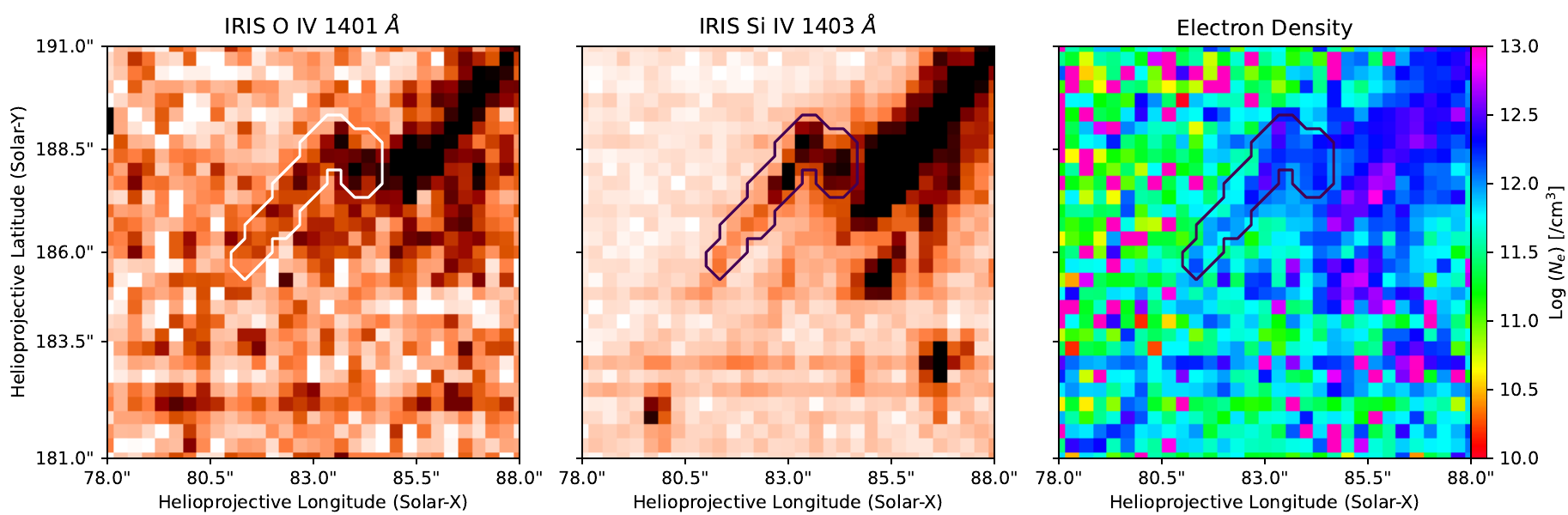}
\caption{JET1 seen in IRIS data. The left and middle panels show the O~{\sc iv} 1401~\AA\ and Si~{\sc iv} 1403~\AA\ intensities. The right panel displays a map of the obtained electron density.}\label{fig:iris_density}
\end{figure*}

In this section, we describe the method for estimating the electron number density ($N_e$) of the emitting plasma using an emissivity ratio method \citep{1994A&ARv...6..123M}. First, raster images of Si~{\sc iv} 1403\,\AA~and the O~{\sc iv} 1401\,\AA~are generated by summing the intensity of the spectrum within the velocity range of $-$30 to $-$10 \kmps, as shown in Fig. \ref{fig:iris_density}. Then we selected pixels of the UV/EUV brightenings at the top of the jet by comparing the slit positions and the slit-jaw animations carefully. We disregard the pixels where the digital number is lower than 50  in the Si~{\sc iv} raster images. Finally, pixels inside the solid line contour in Fig. \ref{fig:iris_density} are selected to estimate the electron density.

To calculate $N_e$ from the intensity ratio of the Si~{\sc iv} and O~{\sc iv} lines, we use the theoretical ratio using the atomic data from CHIANTI v.11. The maximum formation temperature of the peak of the ion fraction is chosen to T=10$^{4.87}$~K and T=10$^{5.19}$~K for Si~{\sc iv} and O~{\sc iv}, respectively. The calculations were done with the photospheric abundances of \citet{asplund_2021} and the ionization fraction for optically thin plasmas of \citet{mazzota_1998} under the assumption of ionization equilibrium, as included in CHIANTI. As a result, $N_e$ is calculated as $10^{12.1\pm 0.2}$\,cm$^{-3}$ from the selected pixels. 

\bibliography{ms_hsyang}





\end{document}